\documentclass[aps,prl,twocolumn,groupedaddress]{revtex4}
\def\p{{_{\bf P}}}
\usepackage{graphicx}

\begin{document}
\preprint{12345678}

\title{Pattern formation in growing sandpiles}
\author{Deepak Dhar}
\email[]{ddhar@theory.tifr.res.in}
\homepage[]{www.theory.tifr.res.in/~ddhar}
\author{Tridib Sadhu}
\email[]{tsadhu@gmail.com}
\homepage[]{www.theory.tifr.res.in/~tridib}
\author{Samarth Chandra}
\email[]{schandra@tifr.res.in}
\homepage[]{www.theory.tifr.res.in/~schandra}
\affiliation{Department of Theoretical Physics, Tata Institute of 
Fundamental Research, $1$ Homi Bhaba Road,  Mumbai 400005 India }
\date{\today}
\begin{abstract}
Adding grains at a single site on a flat substrate in the Abelian 
sandpile models produce beautiful complex patterns. We study in detail 
the pattern produced by adding grains on a two-dimensional square 
lattice with directed edges (each site has two arrows directed inward 
and two outward), starting with a periodic background with half the 
sites occupied.  The size of the pattern formed scales  with the 
number of grains added $N$ as $\sqrt{N}$.  We give exact 
characterization of the asymptotic pattern, in terms of the position
and shape of different features of  the pattern.
\end{abstract}
\maketitle
Many complicated and intricate patterns found in nature can 
be modelled by deterministic dynamics \cite{cross}. In Turing patterns 
\cite{ref2} the final outcome is random due to the randomness in initial 
conditions. In the game of life \cite{ref3}, one can get a very wide 
variety of patterns from  simple deterministic cellular automaton 
evolution rules, depending on the initial condition.

While the real sand, poured at one point on a flat substrate produces a 
rather simple pyramidal shape, much more complex patterns are produced 
in the theoretical models of sandpiles, like the Abelian sandpile model 
(ASM) \cite{dhar06}.  Earlier studies have usually concentrated on 
determining the asymptotic shape of the growing cluster 
\cite{dhar99,redig}. Other special configurations in the model, like the 
identity \cite{identity}, or the stable state produced from special 
unstable states also show complex internal self-similar structures 
\cite{liu}.  The limiting shape has been determined in the related 
rotor-router model, and the model of divisible sandpiles with multiple 
sites of addition \cite{lionel}.

In this paper, we study the asymptotic pattern produced by adding $N$ 
grains of sand at a single site on a two dimensional Abelian sandpile 
model starting from a periodic background, and allowing the system to 
relax.  It is easy to see that the diameter of the pattern grows as 
$\sqrt{N}$. Interestingly, for large $N$, the pattern shows a {\it 
proportionate} growth, with different parts of the pattern all growing 
as $\sqrt{N}$. This is thus different from earlier-studied models of 
growth such as diffusion limited aggregation, Eden model etc. \cite{growth},
where the growth occurs mainly at the surface.

\begin{figure}
 \begin{center}
  \includegraphics[width=4.04cm,angle=90]{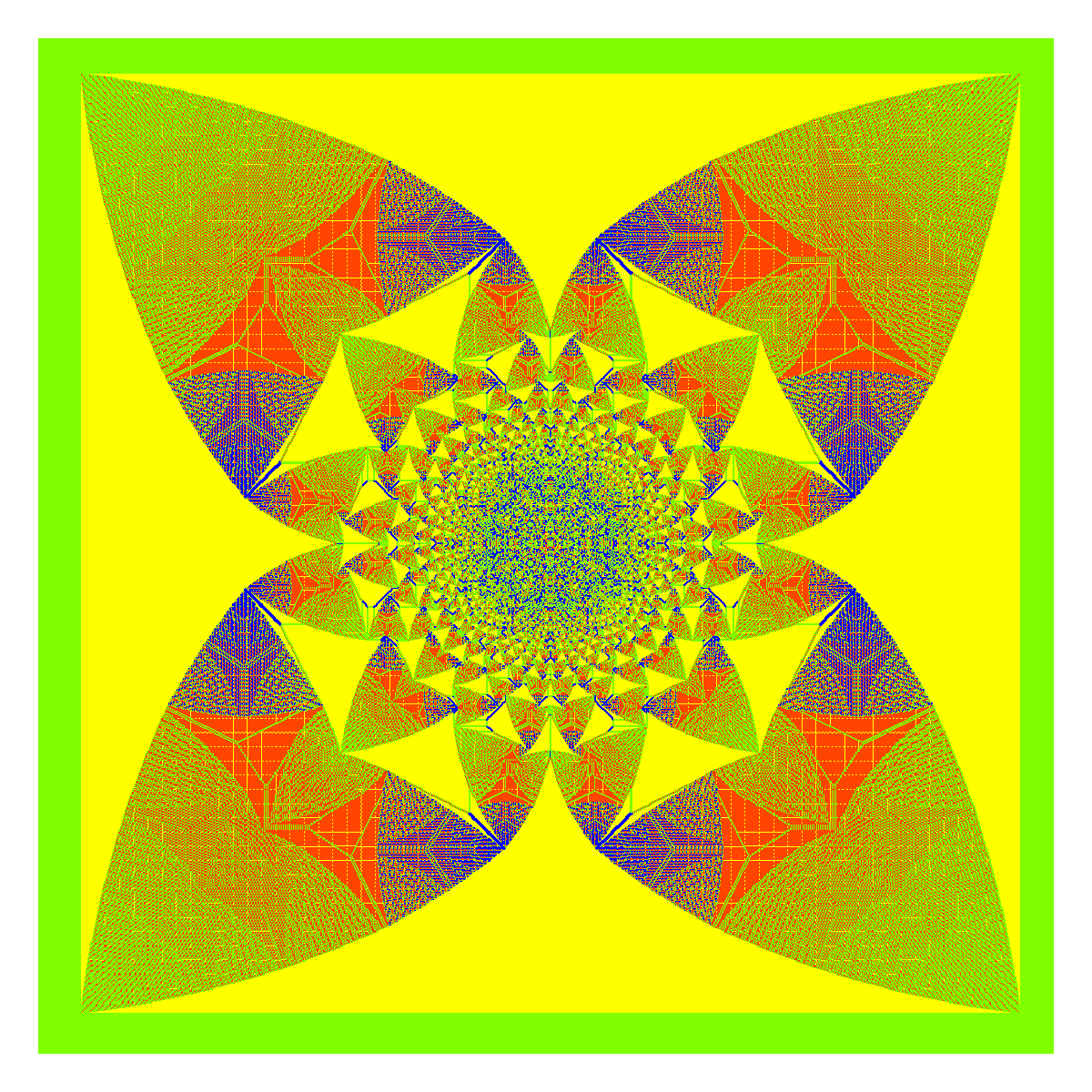}
  \includegraphics[width=4.0cm,angle=90]{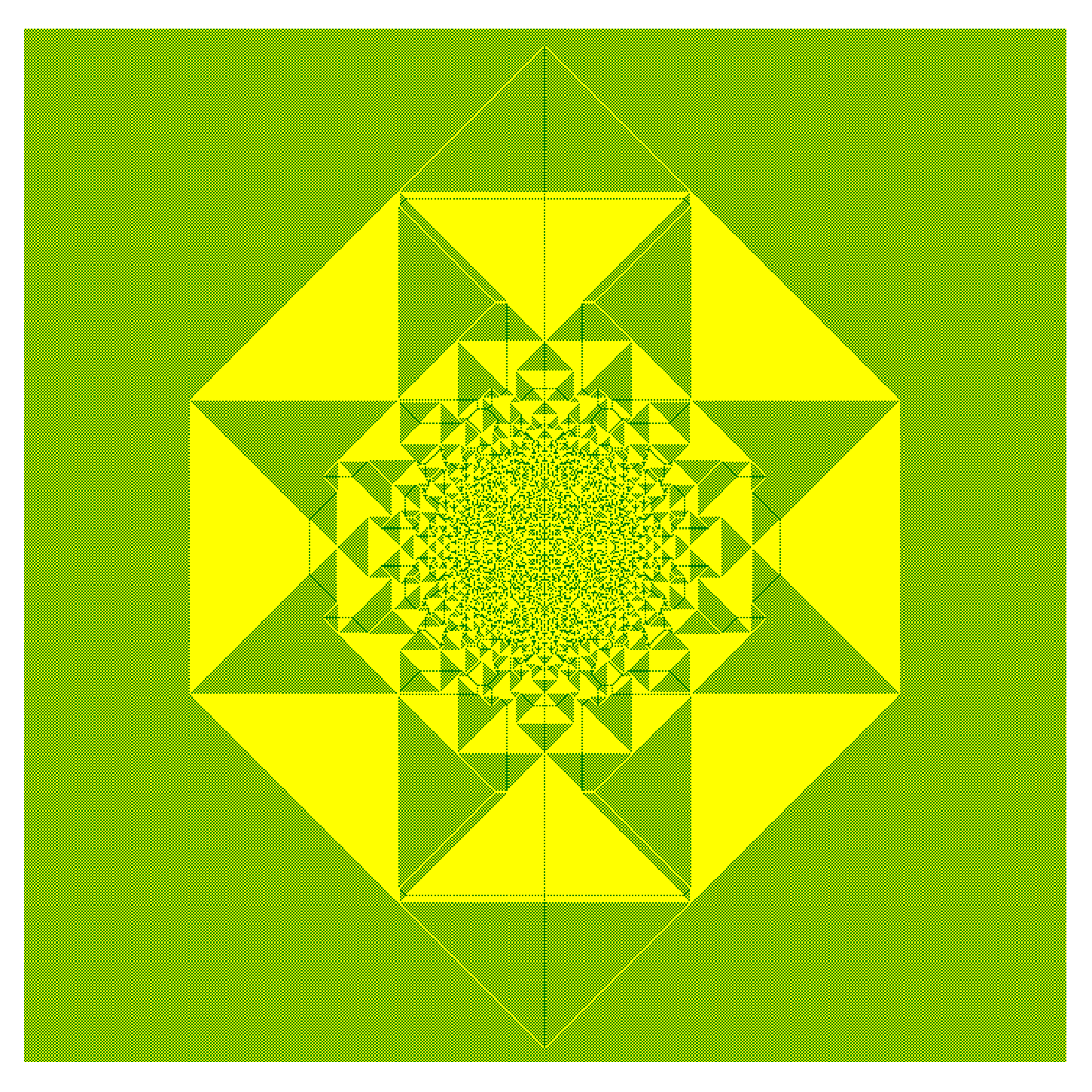}
  \caption{Stable configurations for the Abelian sandpile model obtained 
  by adding particles at one site. $(a)$ Undirected square lattice, initial 
  configurations with all heights $2$, and  $2\times 10^5$ particles 
  added, color code: red=0, blue=1, green=2, yellow=3. $(b)$ The F-lattice of 
  Fig.\ref{lattice}$a$ with initial checkerboard configuration, with 
  $2\times10^5$ particles added, color code: green=0, yellow=1. The apparent
  green regions in the picture represent the patches with checkerboard
  configuration.  }
  \label{fig1}
 \end{center}
\end{figure}

\begin{figure}
 \includegraphics[width=3.5cm,angle=90]{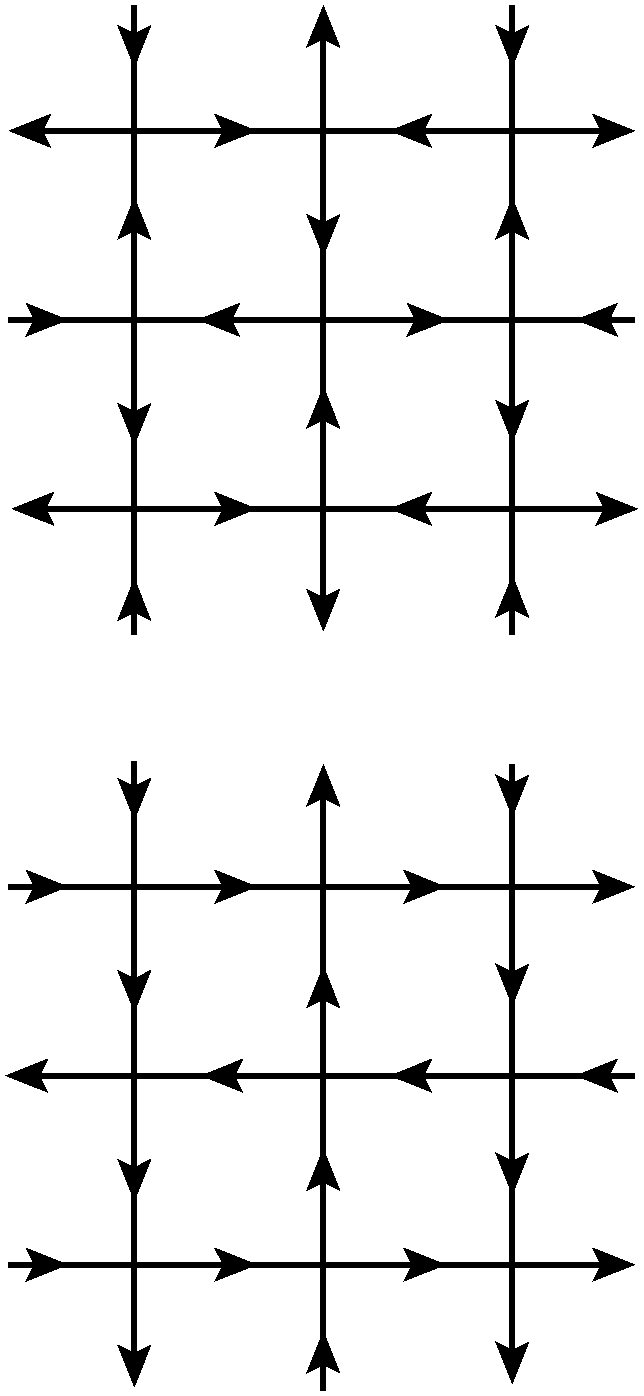}
 \caption{The directed square lattices studied in this paper $(a)$ the  
 F-lattice $(b)$ the Manhattan lattice.}
 \label{lattice}
\end{figure}

The standard square lattice produces a rather complicated pattern 
(Fig.\ref{fig1}$a$), and it has not been possible to characterize 
it so far.  We consider two variations, assigning orientations to 
the edges of the lattice, as shown in Fig.\ref{lattice}$a$ and 
\ref{lattice}$b$. The initial state was chosen to be a periodic checkerboard 
arrangement of sites with heights $0$ and $1$. The asymptotic pattern 
produced in the two cases turns out to be the same, and is shown in Fig 
\ref{fig1}$b$. Taking some qualitative features of the observed pattern
( e.g. only 
two types of patches are present, and they are all $3$- or $4$- sided 
polygons) as input, we show how one can get a complete and {\it 
quantitative} characterization of the pattern.  We show that the pattern 
has exact $8$-fold rotational symmetry, and determine the exact 
coordinates of all the boundaries in the asymptotic pattern.  We discuss 
some other cases, where exactly the same pattern is obtained.

In the two lattices we studied (Fig.\ref{lattice}), each bond of the lattice is 
directed with two in-arrows, and two out-arrows at each vertex. The ASM 
on these is defined by the toppling rule: A site $(x,y)$ is unstable if the 
number of grains at the site $z_{x,y}\ge2$, and then transfers one grain 
each in the direction of its outward arrows. We start with an initial 
configuration in which  $z_{x,y}=1$, for sites with $(x+y)=$ even, and $0$ 
otherwise.

We used a lattice large enough so that no avalanches started from the 
origin reach the boundary.  Using the Abelian property, we add all $N$ 
particles in the beginning, and relax the configuration to get the final 
pattern.  The result of adding $N=2\times 10^5$ particles on the 
F-lattice is shown in Fig.\ref{fig1}$b$. The pattern formed on the Manhattan 
lattice is indistinguishable at large scales. The pattern is identical 
to Fig.\ref{fig1}$b$, except that the thin lines of $1$'s forming two triangles 
outside the octagon are rotated by $45^{\circ}$ in the 
Manhattan case.  Since the lattices are quite different, this is quite 
intriguing.
   
We start by setting up some general theoretical framework, which is 
independent of the details of the particular lattices studied. Formally, we 
can characterize the asymptotic pattern in terms of the rescaled 
coordinates, $\xi=x/\sqrt{N}$, $\eta=y/\sqrt{N}$ and the density function 
$\rho(\xi,\eta)$ which gives the local density of grains 
in the pattern in a small rectangle of size $\Delta\xi$, $\Delta\eta$ 
about the point $(\xi,\eta)$, with $N^{-1/2} \ll\Delta\xi, \Delta 
\eta\ll1$.

Equivalently, we can describe the asymptotic pattern in term of the
rescaled toppling function $\phi(\xi,\eta)$. Let $T_N(x,y)$ be the numbers of
toppling  at site $(x,y)$ when $N$ particles are added at
the origin, and the configuration is relaxed. We define  
\begin{equation}
\phi(\xi,\eta)= \lim_{N\rightarrow \infty}\frac{1}{2N}T_N(\lfloor
\sqrt{N}\xi \rfloor, \lfloor \sqrt{N}\eta \rfloor)
\end{equation}
where floor function $\lfloor x \rfloor$ is the largest integer less 
than $x$. From the conservation of sand grains, it is easily seen that 
$\phi(\xi,\eta)$ is related to the density function $\rho(\xi,\eta)$ by
\begin{equation}
(\frac{\delta^2}{\delta\xi^2} + \frac{\delta^2}{\delta\eta^2}
)\phi(\xi,\eta)=\Delta\rho(\xi,\eta) - \delta(\xi)\delta(\eta)
\label{poisson}
\end{equation}
where excess density $\Delta\rho(\xi,\eta)$ is the difference between 
$\rho(\xi,\eta)$ and the initial density $\rho_0(\xi,\eta)$. 

It was already noted \cite{ostojic} that for $N$ large $\rho(\xi,\eta)$ 
tends to a nontrivial limit, and the asymptotic pattern is made of 
distinct regions, called `patches'. Typically inside a patch the heights
are periodic in space, and there are few 
defect-lines, which move with $N$, but do not change the macroscopic 
density $\rho(\xi,\eta)$.  Then, the coarse grained function 
$\rho(\xi,\eta)$ takes constant rational value in each patch.
Also in each patch of constant $\Delta\rho(\xi,\eta)$, $\phi(\xi,\eta)$ 
is a quadratic function, and was first noted in \cite{ostojic}. We
indicate the proof here. For all patches the function $\phi(\xi,\eta)$ is
Taylor expandable around any point inside the patch. Consider any term
of order $\ge 3$ in the expansion, for example the term
$\sim(\Delta\xi)^3$. This can only arise due to a term $\sim (\Delta
x)^3/\sqrt{N}$ in $T(x,y)$. Then the integer function $T(x,y)$ will 
change discontinuously at intervals of $\Delta x \sim \mathcal{O}(N^{1/6})$
leading to infinitely many defect-lines in the asymptotic pattern. However
there are no such feature in Fig.\ref{fig1}$a$ or Fig.\ref{fig1}$b$.
Therefore inside a patch of constant $\Delta\rho(\xi,\eta)$,
$\phi(\xi,\eta)$ can at most be quadratic in $\xi$ and $\eta$, and in
each periodic patch, the toppling function $T(x,y)$ is sum of two terms: a 
part that is a simple quadratic function of $x$ and $y$, and a periodic 
part. The periodic part averages to zero, and does not contribute to the 
coarse-grained function $\phi(\xi,\eta)$. In some patterns, there are 
regions of finite fractional area which show aperiodic height 
patterns. In these regions $\phi(\xi,\eta)$ is not quadratic
and are harder to characterize.

\begin{figure}
 \includegraphics[scale=0.25]{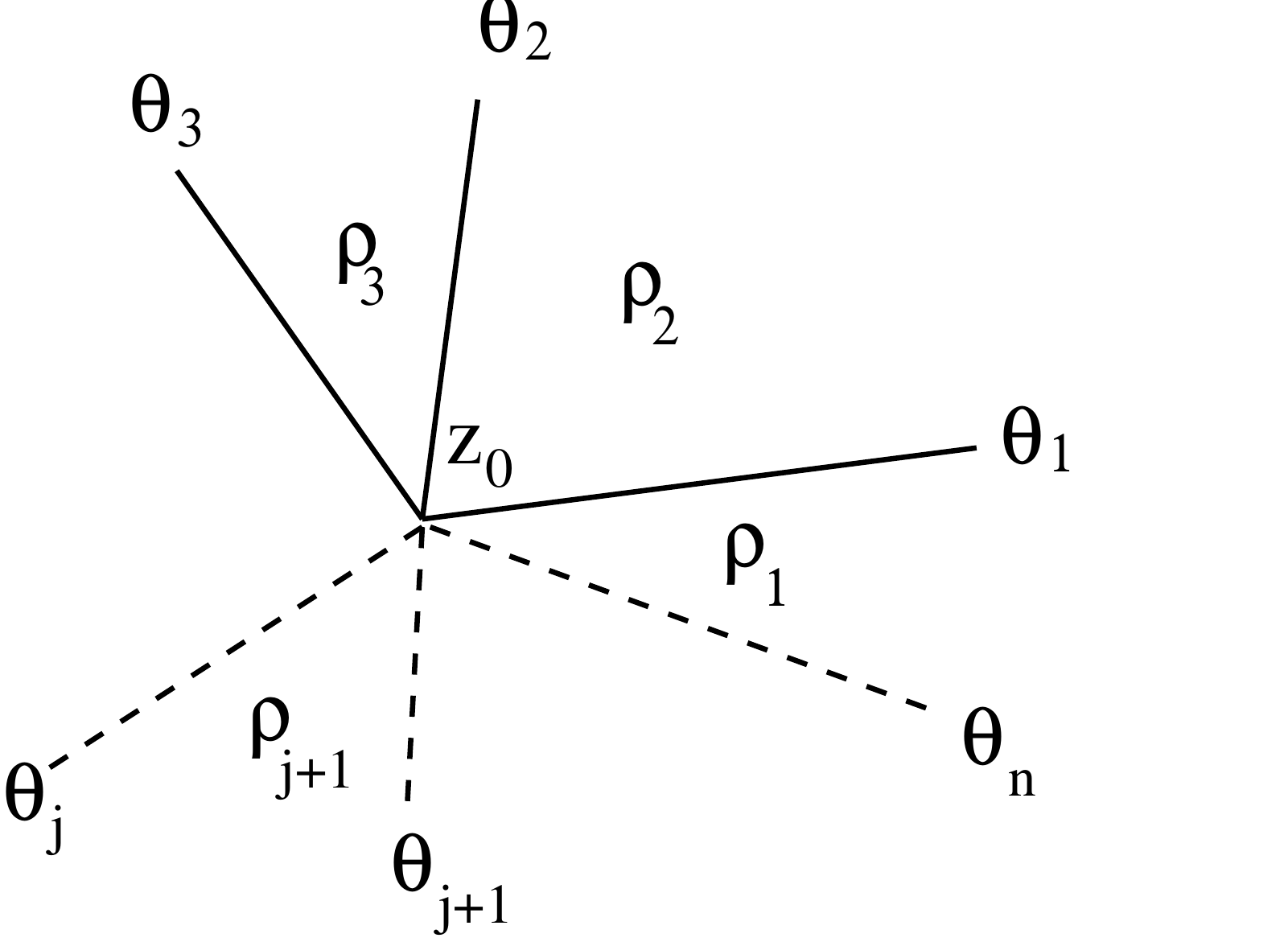}
 \includegraphics[scale=0.05]{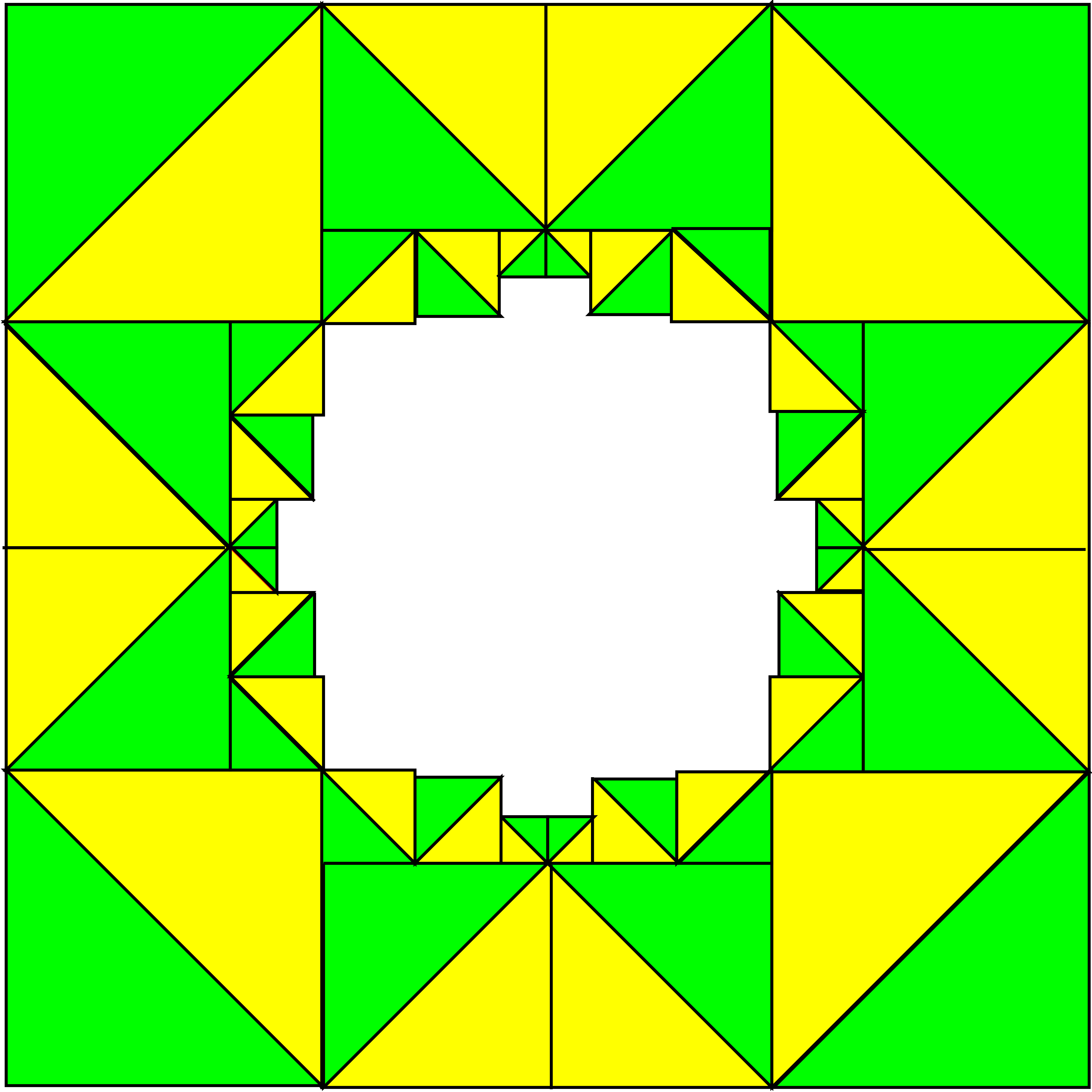}
 \caption{$(a)$ $n$ different periodic patches of density
 $\rho_1$,$...$,$\rho_n$ meeting at point $z_0$. $(b)$ The pattern in 
 Fig.\ref{fig1}$b$ is obtainable by putting together square tiles of 
 different sizes. Each of the tiles is divided into two halves of 
 different density.}
 \label{tile}
\end{figure}

Now consider two neighboring periodic patches ${\bf P}$ and ${\bf P'}$ 
with mean densities $\rho$ and $\rho'$ respectively.  Let the quadratic 
toppling function be $Q(\xi,\eta)$ and $Q'(\xi,\eta)$ in these 
patches. Then the boundary between the patches is given by the equation 
$Q(\xi,\eta) = Q'(\xi,\eta)$. As the derivatives of $\phi$ are also 
continuous across the boundary, the boundary between two periodic 
patches must be a straight line, and
\begin{equation} 
Q'(\xi,\eta) = Q(\xi,\eta)+\frac{1}{2}(\rho'-\rho) l_{\perp}^2  
\label{continuity}
\end{equation} 
where $l_\perp$ is the perpendicular distance of $(\xi,\eta)$ from the 
boundary. We can start with a periodic patch $P$, and go to another 
patch $P'$ by more than one path. Since the final quadratic function at 
$P'$ should be the same whichever path we take, this imposes consistency 
conditions which restricts the allowed values of slopes of boundaries. 
Consider a point $z_0$ where $n$ periodic patches meet, with $n > 2$
(Fig.\ref{tile}$a$). If the $j$th boundary at this point makes an angle 
$\theta_j$ with the $x$-axis, and the density of the patch in the wedge 
$\theta_j \le \theta \le \theta_{j+1}$ is $\rho_{j+1}$ (Fig.\ref{tile}$a$) 
then using Eq.\ref{continuity} repeatedly for all $n$ patches around $z_0$
we get that the following equation must be true for all $\theta$:
\begin{equation}
\sum_{j=1}^n (\rho_{j+1} -\rho_{j}) \sin^2( \theta - \theta_j) =0,
\end{equation}
with $\rho_{n+1}=\rho_n$. This is equivalent to the  condition:
\begin{equation}
\sum_{j=1}^n  (\rho_{j+1} -\rho_{j}) e^{ 2 i \theta_j}  =0
\end{equation}
For $n=3$, with $\rho_1 \neq \rho_2 \neq \rho_3$, this equation has
only trivial solutions with $\theta_{j}$ equal to $0$ or $\pi$ for all
$j$. Hence, only  $n \geq 4$ are allowed.

We now discuss how the exact function $\rho(\xi,\eta)$ can be determined 
for our problem. We note that in Fig.\ref{fig1}$b$, there are no aperiodic patches, 
only two types of periodic patches, where $\rho(\xi,\eta)$ only take 
values $1$ or $1/2$. Also, the slopes of the boundaries between patches 
only take values $0$, $\pm1$, $\infty$. The patches are typically dart 
shaped quadrilaterals, and some triangles (which may be considered as 
degenerate quadrilaterals with one side of length zero). These 
simplifications, not present in Fig.\ref{fig1}$a$, make possible a full 
characterization of the pattern in Fig.\ref{fig1}$b$.
 
Given that there are only these two types of patches, we only need to look 
for possible patterns where $\Delta \rho$ takes piecewise constant 
values $1/2$ or $0$. From Eq.(\ref{poisson}), we see that we can think 
of $\phi(\xi,\eta)$ as the potential produced by a point charge at the 
origin, and a charge cloud with areal density $-\Delta\rho(\xi,\eta)$, 
with total charge zero. The basic principle which selects the actual 
stable pattern out of many is a version of the principle of minimum
dissipation: {\it It is a stable state reached by minimum number of
toppling.} (This follows immediately from the 
toppling rules, where no toppling occurs unless forced).

The requirement that $\phi(\xi,\eta)$ be exactly zero, in the region 
outside the pattern, implies that all the multipole moments of the 
charge distribution $\Delta\rho(\xi,\eta)$ are exactly zero. We show below 
that the conditions that $\Delta \rho$ takes only two values, the potential
function is exactly quadratic within a periodic patch, and the 
slopes of the boundaries are only $0,\pm1,\infty$, fix the allowed pattern 
uniquely.

We start by determining the exact asymptotic size of the pattern. We
note from Fig.\ref{fig1}$b$ that the boundary of the pattern is an octagon (
we shall prove later that this is a regular octagon ). In fact there
are four lines of $1$'s outside the octagon. But these has zero areal
density in the limit $N\rightarrow \infty$, and do not contribute to 
$\rho(\xi,\eta)$. We will ignore these  in
the following discussion.

Let $B$ be the minimum boundary square containing all ($\xi$, $\eta$)
that have a non-zero charge density $\rho(\xi, \eta)$. We
observe that $B$ can be considered as a union of
disjoint smaller squares, each of which is divided by diagonal into two 
parts where $\Delta \rho(\xi,\eta)$ takes values $1/2$ and $0$
[Fig.\ref{tile}$b$]. This is seen to be true for the
outer layer patches. Towards the center, the squares are not so well
resolved. Assuming that this construction remains true all the way to the
center, in the limit of large $N$, the mean density of the negative 
charge in the bounding square $=1/4$. Given that the total amount of 
negative charge is $-1$, the area of the bounding square should be $4$. 
Hence we conclude that the equation of the boundary of the minimum 
bounding square are
\begin{equation}
|\xi|=1, \rm{~~~~~~} |\eta|=1
\end{equation}
Let $N_b$ be the minimum number of particles that have to be added so
that at least one site at $y=b$ topples. We find that for $b=10$,
$50$, $100$, and $300$, $\sqrt{N_b}=10.770$, $49.436$, $98.894$ and $297.798$.
Clearly the boundary distance $b$ tends to $\sqrt{N}$ for large $N$.  

\begin{figure}
  \begin{center}
    \begin{tabular}{cc}
    \resizebox{31mm}{!}{\includegraphics[height=40mm, 
    width=40mm]{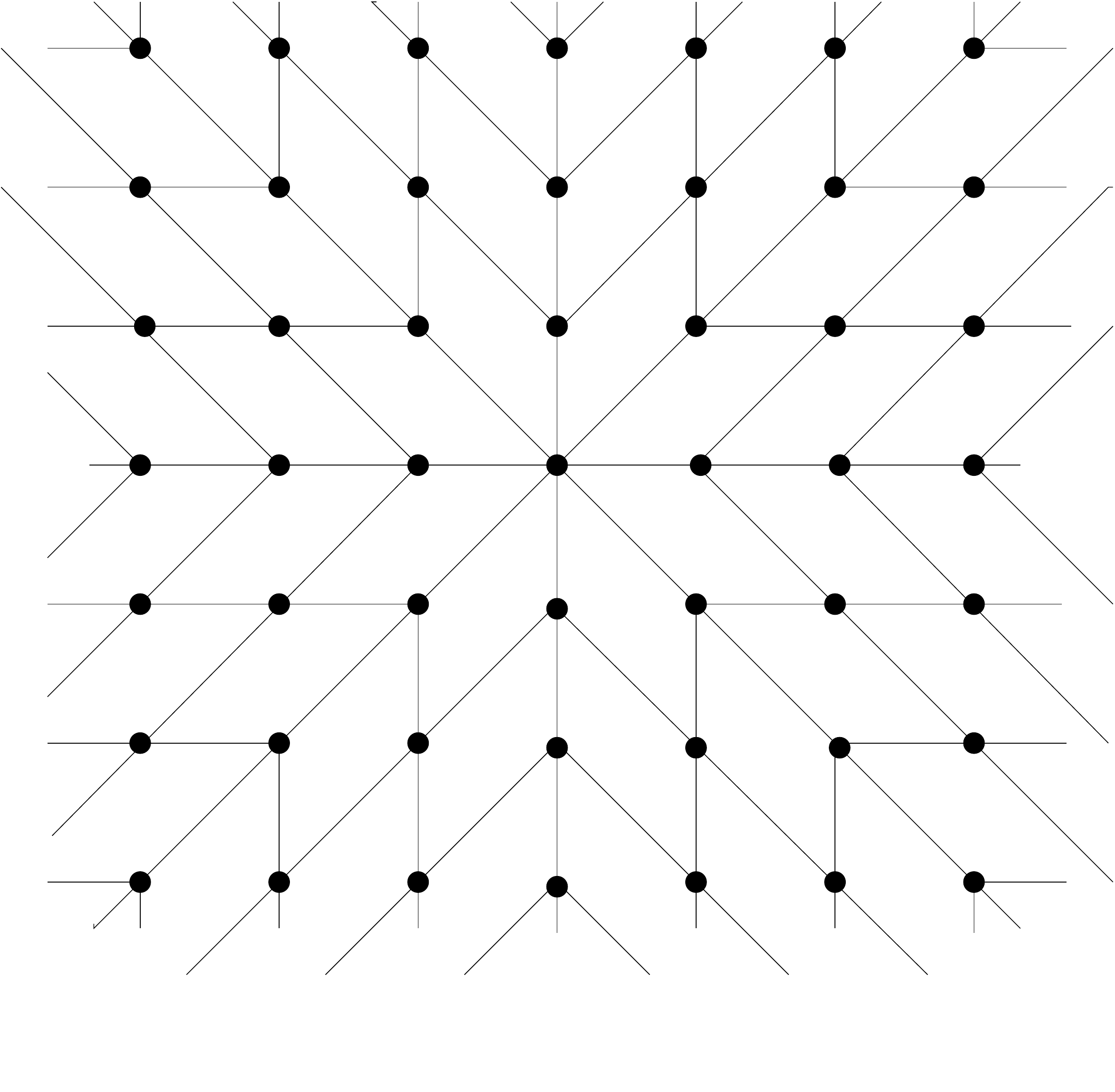}} &    
    \resizebox{32mm}{!}{\includegraphics[height=40mm, 
    width=40mm]{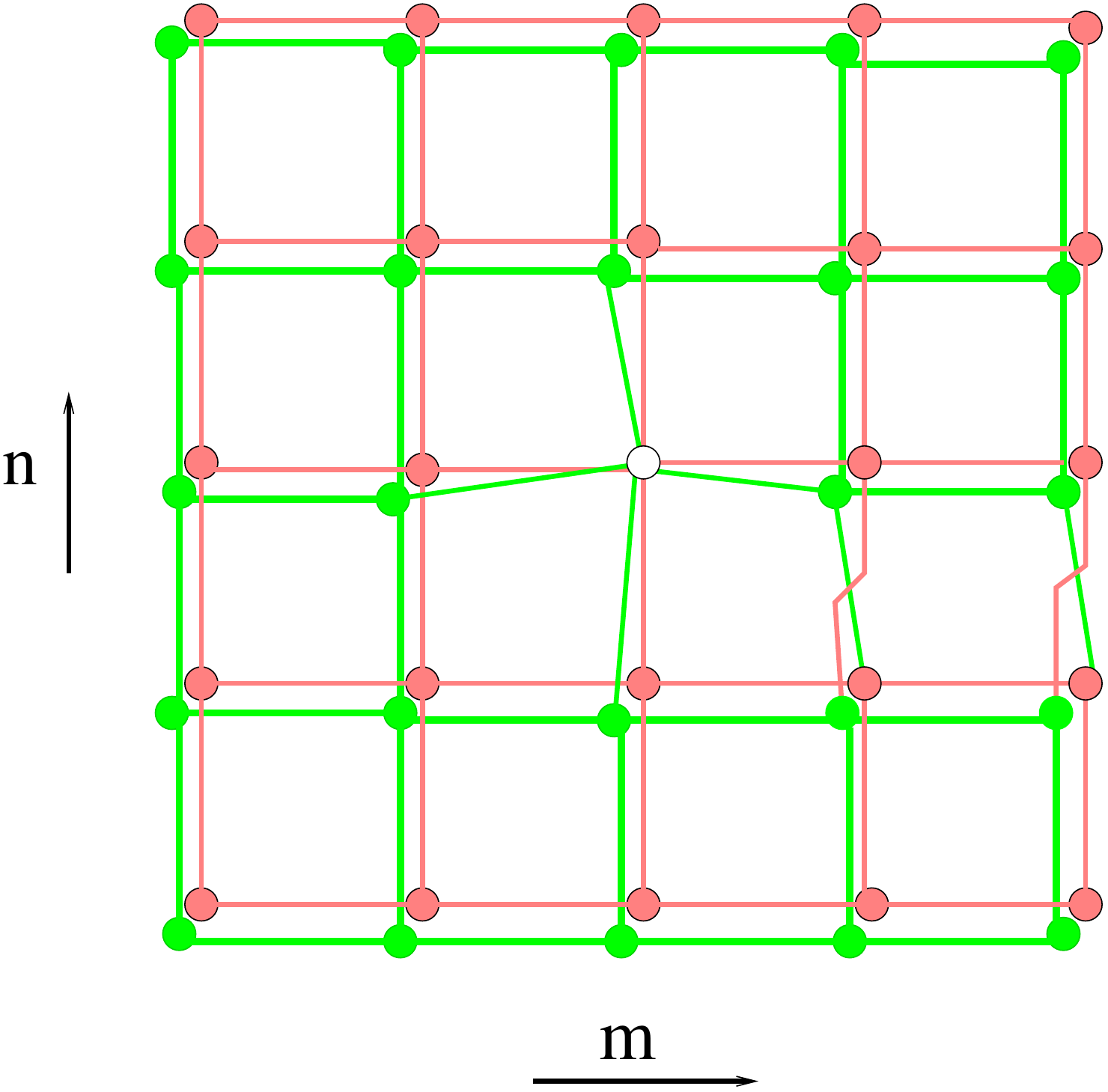}} 
    \end{tabular}
    \caption{Two representations of the adjacency graph of the pattern. 
    Here the vertices are the patches, and the edges connect the adjacent 
    patches.
    $(a)$ Representation as a planar graph $(b)$ as a graph of wedge of angle 
    $4 \pi$ formed by glueing together the eight quadrant graphs at the
    origin.}
    \label{fig5}
  \end{center}
\end{figure}

We now describe the topological structure of the pattern. This is 
characterized by its adjacency graph [Fig.\ref{fig5}$a$], where each vertex 
denotes a patch, and a bond between the vertices is drawn if the 
vertices share a common boundary. It is convenient to think of the 
triangular patches in the pattern as degenerate quadrilaterals, with one 
side of length zero. Then we see that the adjacency graph is planar with 
each vertex of degree four, except a single vertex of coordination 
number eight corresponding to the exterior of the pattern. The graph has 
the structure of a square lattice wedge, with wedge angle $4\pi$. The 
square lattice structure of the adjacency graph is seen most directly by 
applying a $z' = 1/z^2$ transformation to the picture (used earlier in 
\cite{ostojic}), where $z=\xi+ i \eta$, and view it in the complex 
$z'$-plane. Thus, one can equivalently represent the graph as a square 
grid on a Riemann surface of two sheets (fig.\ref{fig5}$b$).

We now use the qualitative information obtained from the adjacency 
matrix of the observed pattern, to obtain quantitative prediction of the 
exact coordinates of all the patches. Consider an arbitrary patch ${\bf P}$, 
having an excess density $1/2$. The potential function in the patch is a 
quadratic function of $(\xi,\eta)$ and we parametrize it as
\begin{eqnarray}
\phi_\p(\xi,\eta) &=& \frac{1}{8}(m_\p+1)\xi^2 + \frac{1}{4}n_\p\xi\eta
+\frac{1}{8}(1-m_\p)\eta_p^2 \nonumber \\
&&+ d_\p\xi + e_\p \eta + f_\p
\end{eqnarray}
The potential function in a patch ${\bf P}$ having zero excess density 
will be parametrized as
\begin{equation}
\phi_\p(\xi,\eta)=\frac{1}{8}m_\p(\xi^2-\eta^2) + \frac{1}{4}n_\p\xi\eta
 + d_\p\xi + e_\p\eta + f_\p
\end{equation}
Now consider two neighboring patches ${\bf P}$ and ${\bf P'}$ with   
excess densities  $1/2$ and $0$ respectively. Then using the matching 
condition Eq.(3), it  is easy to show that if the boundary between  them 
is a horizontal line $\eta =\eta_{\p}$, we must have 
\begin{eqnarray}
m_{\p'} &=& m_\p+1 \rm{,~~~}n_{\p'} = n_\p \rm{,~~~} d_{\p'} = d_\p
\nonumber \\
e_{\p'} &=& e_\p + \eta_\p/2 \rm{,~~~} f_{\p'} = f_\p - \eta_\p^2/4
\label{a2}
\end{eqnarray}
Similar calculation for other boundaries show that across a vertical
boundary, going from a patch of higher density to the one of lower
density, we have $\Delta m_\p=-1$
and $\Delta n_\p=0$. Across a boundary with slope $\pm1$, $\Delta
m_\p=0$, and $\Delta n_\p = \pm 1$.

In the outermost patch, clearly $\phi(\xi,\eta)=0$, and for this patch
both $m$ and $n$ are zero. It follows that all $m_\p$ and $n_\p$ take
integer values. In the following, we denote a patch by integers
$(m,n)$, and write the corresponding coefficients $d_\p$, $e_\p$, and
$f_\p$ as $d_{m,n}$, $e_{m,n}$ and $f_{m,n}$. With this convention,
the matching conditions in Eq.(\ref{a2}) can be rewritten as 
\begin{equation}
d_{m+1,n}=d_{m,n} \rm{,~} e_{m+1,n}-e_{m,n}=\eta_{m,n}/2,
(m+n)\rm{~odd}
\label{10}
\end{equation}
Using similar matching conditions for the boundary
of patch $(m$, $n)$ with slope $\pm1$, we get the conditions
\begin{eqnarray}
d_{m,n+1}-d_{m,n} &=& e_{m,n} -e_{m,n+1},(m+n)\rm{~odd}
\nonumber \\
d_{m,n-1}-d_{m,n} &=& e_{m,n-1} - e_{m,n},(m+n)\rm{~odd}
\label{11}
\end{eqnarray}
We can eliminate the variables $d_{m,n}$ and $e_{m,n}$ with $(m+n)$
even using Eq.(10) and Eq.(11). Then the equations become 
\begin{eqnarray}
e_{m+2,n}-e_{m,n} &=& \eta_{m,n}/2 \label{a6}\\
d_{m-2,n}-d_{m,n} &=& \xi_{m,n}/2 \label{a7}\\
d_{m-1,n-1}-d_{m,n} &=& e_{m+1,n-1}-e_{m,n} \label{a8}\\
d_{m-1,n+1}-d_{m,n} &=& -[e_{m+1,n+1}-e_{m,n}] \label{a9} 
\end{eqnarray}

It is convenient to introduce the complex variables $z = \xi + i \eta$,
$ M = m + i n$ and $D = d + i e$. In these variables we can write
Eq.($7$) as
\begin{equation}
\phi(z) =  \frac{1}{8} z \bar{z} + \frac{1}{8} Re[ z^2 \bar{M} + \bar{D} 
z] +f 
\end{equation}

Under a rotation of axes by an angle $\theta$, $z \rightarrow z'= z e^{i 
\theta}$, the requirement that  $\phi$ is invariant is satisfied if we 
have
\begin{equation}
M' = M e^{2 i \theta};~~~~ D' = D e^{i \theta}
\end{equation}

On the $(m,n)$ lattice, with $(m + n)$ odd, the natural basis vectors are 
$(1,1)$ and $(1,-1)$. Let us call these $\alpha$ and $\beta$. We define 
the finite difference operators  $\Delta_{\pm \alpha}$ and $\Delta_{\pm 
\beta}$ by 
\begin{eqnarray}
\Delta_{\pm \alpha} f(z) = f( z \pm \alpha) - f(z)\nonumber\\
\Delta_{\pm \beta} f(z) = f(z \pm \beta) -f(z)
\end{eqnarray}
Then the equations ($14$-$15$) can be written as
\begin{eqnarray}
\Delta_{-\alpha} d = \Delta_{\beta} e \nonumber \\
\Delta_{-\beta} d = -\Delta_{\alpha} e
\end{eqnarray}

These equations are the discrete analog of the familiar Cauchy-Riemann
conditions connecting the partial derivatives of real and imaginary 
parts of an analytic function where the role of the analytic function is 
played by   $ D = d + i e$.   

From Eq.$(14)$ and Eq.($15$), it is easy to deduce that $D$   
satisfies  the  discrete Laplace's equation
\begin{equation}
[\Delta_{\alpha} \Delta_{-\alpha} +\Delta_{\beta} \Delta_{-\beta}] D =0
\label{a10}
\end{equation}

If $m$ and $n$ are large, the corresponding patch is near the origin 
($|\xi|+|\eta|$ is small), and where the leading behavior of 
$\phi(\xi,\eta)$ is given by 
$\tilde{\phi}(\xi,\eta) \sim -\frac{1}{4\pi}\log(\xi^2+\eta^2)$.
Consider a point $z_0$, such that at $z_0$
\begin{equation}
\partial^2 \tilde{\phi}/\partial \xi^2 \approx m/4;~~~~~
\partial^2 \tilde{\phi}/\partial \xi \partial \eta \approx n/4,
\end{equation}
Then, $z_0$ would be expected to lie in the patch labeled by $(m,n)$.
This gives $z_0 \approx \pm ( \pi \bar{M}/2)^{-1/2}$. Then, setting
$\partial \tilde{\phi}/\partial z$ equal to $\bar{D}/2$ gives us
\begin{equation}
D_{m,n} \simeq \pm \frac{1}{\sqrt{2\pi}} \sqrt{m + i n}
\label{a12}
\end{equation}
The equation (\ref{a10}), subjected to the behavior at large
$|m|+|n|$ given by Eq.(\ref{a12}) on the $4 \pi$-wedge graph 
(for each value of $(m,n)$, $D_{m,n}$ has two values) has an unique
solution. Clearly the solution has eight fold rotational symmetry
about the origin in the $(m, n)$ space. This implies that 
\begin{equation}
D_{-n,m}=i^{1/2} D_{m,n}; \rm{~for~all~} (m,n). 
\end{equation}
Given $D_{m,n}$, its real and imaginary parts determine $d_{m,n}$ and
$e_{m,n}$, and using Eq.(\ref{a6}, \ref{a7}) we determine the exact
positions of all the patch corners.
The exact eight-fold rotational symmetry of the adjacency graph of the 
pattern, and the fact that $D$ satisfies Eq.($20$) on the adjacency graph 
together imply the eight-fold rotational symmetry of all the distances 
in the pattern. 

We have not been able to find a closed-form formula for $D_{m,n}$.
But the system of coupled linear equations (\ref{a10}) can be determined 
numerically to very good precision by
solving it on a finite grid $-L \le m$, $n\le L$ with the condition in
Eq.(\ref{a12}) imposed exactly at the boundary. We determined $d_{m,n}$ 
and $e_{m,n}$ numerically for $L=100,200,400$, and extrapolated our 
results for $L\rightarrow\infty$. We find $d_{1,0}= 0.5000 $ and 
$d_{2,1} = 0.6464$, in perfect agreement with the exact  theoretical 
values $1/2$ and $1- 1/2\sqrt{2}$ respectively.

Our arguments above can be extended to other two dimensional lattices, 
so long as there are only two allowed values of $\Delta \rho$.  While this is not clear why, 
this seems to happen for the Manhattan lattice (Fig.\ref{lattice}$b$), for initial 
density $1/2$.  Also, this happens on the F-lattice, with a periodic 
background pattern with initial density 5/8 [$z_{i,j}=1$ if $i+j$ 
even, or $(i,j)$ congruent to $(0$, $1)$ or $(2$, $3)$ mod $4$].  In some other 
cases, like the F-lattice, with initially all sites empty, the pattern 
is very similar, but there are some non periodic patches in the outermost 
ring.  Since  the behavior of 
$\phi(\xi,\eta)$ in such patches is not known, the equations for 
$D_{m,n}$ do not close in this case. 
\begin{acknowledgments} 
We thank L. Levine for very useful discussions. The special features of 
growth pattern studied here were noted first in numerical studies by Mr. 
Subhendu B. Singha.  DD would like to thank J. P. Eckmann for getting 
him interested in this problem, and B. Nienhuis for discussions. 
\end{acknowledgments}

\end{document}